\documentstyle[12pt,epsf]{article}

\begin{document}
\baselineskip = 1.5 \baselineskip

\title {A Variational Approach to Nonlocal Exciton-Phonon Coupling}

\author{
Yang Zhao\\Condensed Matter Section\\International Center for Theoretical Physics,\\P.O. Box 586, 34100 Trieste, Italy
\and
David W. Brown\\Institute for Nonlinear Science,\\University of
California, San Diego, La Jolla, CA 92093-0402
\and
Katja Lindenberg\\Department of Chemistry,\\University of California,
San Diego, La Jolla, CA 92093-0340
}

\date{\today}

\maketitle

\begin{abstract}
In this paper we apply variational energy band theory to a form of the Holstein Hamiltonian in which the influence of lattice vibrations (optical phonons) on both local site energies (local coupling) and transfers of electronic excitations between neighboring sites (nonlocal coupling) is taken into account.
A flexible spanning set of orthonormal eigenfunctions of the joint exciton-phonon crystal momentum is used to arrive at a variational estimate (bound) of the ground state energy for every value of the joint crystal momentum, yielding a variational estimate of the lowest polaron energy band across the entire Brillouin zone, as well as the complete set of polaron Bloch functions associated with this band.
The variation is implemented numerically, avoiding restrictive assumptions that have limited the scope of previous assaults on the same and similar problems.
Polaron energy bands and the structure of the associated Bloch states are studied at general points in the three-dimensional parameter space of the model Hamiltonian (electronic tunneling, local coupling, nonlocal coupling), though our principal emphasis lay in under-studied area of nonlocal coupling and its interplay with electronic tunneling; a phase diagram summarizing the latter is presented.
The common notion of a "self-trapping transition" is addressed and generalized.
\end{abstract}

\pagebreak

\section{Introduction}

In a previous paper \cite{Zhao94a}, we undertook a numerical implementation of methods proposed by Munn and Silbey \cite{Munn85} for the determination of polaron properties in the presence of simultaneous local and nonlocal exciton-phonon coupling, local coupling being defined as a nontrivial dependence of the exciton site energies on lattice coordinates and nonlocal coupling as a nontrivial dependence of the exciton transfer integral on lattice coordinates.
Those methods were essentially perturbative, with the coefficients of a canonical transformation being fixed so as to limit the secular growth with the temperature of the perturbation remaining after transformation.
The present work is motivated by several limitations of those methods that were found in the course of our prior work \cite{Zhao94a}:
1) Since the Munn-Silbey method was motivated by an interest in controlling perturbations at high temperatures, one should not necessarily expect the method to produce the best description of polaron states at low temperatures.
2) Since the method is perturbative, it is reasonable to expect that the description of the polaron ground state so obtained could be improved.
3) Since the self-consistency equations on which the method is based do not involve the exciton transfer integral, it is questionable whether the method can be relied upon beyond the very narrow band regime.
4) Since the results of our numerical investigation showed some significant differences with analytical results based on the same underlying methods, as well as with some other recent analyses
\cite{Bryksin91}-\cite{Capone96},
it would appear desirable to have corroboration of our results by independent methods.

The central findings of our prior work \cite{Zhao94a}, that polaron binding energies may be much larger than previously thought, and polaron bands much narrower, underscores the importance of independent corroboration, since such findings are of central importance to understanding the influence of nonlocal exciton-phonon coupling on the nature of polaron states and of polaron transport.

Simultaneous local and nonlocal coupling appears to be particularly important in the characterization of solid-state excimers, where a variety of experimental and theoretical considerations suggest that a strong dependence of electronic tunneling upon certain coordinated distortions of neighboring molecules (nonlocal exction-phonon coupling) is crucial to the formation of excited bound states
\cite{Bryksin91, Sewell63, Tanaka63, Murrell64, Song69, Tanaka74, Umehara79, Toyozawa80, Fischer83, Umehara83, Port84, Walker85, Sumi89, Wu90a, Wu90b, Wu91, Schopka91, Weiss96, Wu93, Zhao94b, Silinsh94}.

In this paper we approach the problem of simultaneous local and nonlocal exciton-phonon coupling by variational methods.

Our central interest in this paper is in the polaron energy band, computed as
\begin{equation}
E^\kappa = \langle \Psi ( \kappa ) | \hat{H} | \Psi ( \kappa ) \rangle
\end{equation}
wherein $\kappa$ is the total crystal momentum label, $| \Psi ( \kappa ) \rangle$ is an appropriately normalized delocalized trial state, and $\hat{H}$ is the system Hamiltonian.
It should be noted that the trial states we use are eigenfunctions of the appropriate total momentum operator, so that variations for distinct $\kappa$ are independent, and the set of $E^\kappa$ so produced constitute a variational estimate (upper bound) for the polaron energy band \cite{Lee53,Toyozawa61}.

As our system Hamiltonian, we choose perhaps the simplest one embracing exciton tunneling and simultaneous local and nonlocal exciton-phonon coupling, a slight generalzation of the traditional Holstein Hamiltonian \cite{Holstein59i,Holstein59ii}.
\begin{equation}
\hat{H} = \hat{H}^{ex} + \hat{H}^{ph} + \hat{H}^{ex-ph}
\end{equation}
\begin{equation}
\hat{H}^{ex} = - J \sum_n a_n^{\dagger} ( a_{n+1} + a_{n-1} )
\end{equation}
\begin{equation}
\hat{H}^{ph} = \omega \sum_n b_n^{\dagger} b_n
\end{equation}
\begin{equation}
\hat{H}^{ex-ph} = g \omega \sum_n a_n^{\dagger} a_n ( b_n^{\dagger} + b_n ) + \frac {\phi} {2} \omega \sum_n ( a_n^{\dagger} a_{n-1} + a_{n-1}^{\dagger} a_n - a_{n+1}^{\dagger} a_n - a_n^{\dagger} a_{n+1} ) ( b_n^{\dagger} + b_n )
\end{equation}
in which $a_n^\dagger$ creates an exciton in the rigid-lattice Wannier state at site $n$, and $b_n^\dagger$ creates a quantum of vibrational energy in the Einstein oscillator at site $n$.
Both excitons and phonons are treated as bosons.
The Einstein frequency is given by $\omega$, $J$ is the exciton transfer integral between nearest neighbor sites, $g$ is the local coupling strength, and $\phi$ is the nonlocal coupling strength characterizing phonon assisted transfers between nearest neighbor sites.
All of the methods used in this paper can be applied as well to common generalizations of this Hamiltonian, involving, for example, phonon dispersion, exciton transfers beyond nearest neighbors, and/or different exciton-phonon coupling geometry.

In Ref~\cite{Zhao94a} we considered both {\it antisymmetric} nonlocal coupling, as above, and {\it symmetric} nonlocal coupling, in which all nonlocal coupling terms would be of the same algebraic sign.
These two kinds of coupling represent different physical circumstances.
Antisymmetric coupling, for example, would be appropriate to the description of certain librations that promote exciton transfers between neighboring molecules when these molecules tilt toward each other, effectively decreasing the gap through which tunneling must take place.
This has the consequence that tunneling between a molecule and its neighbor to the right (for example) is promoted (and tunneling on the left inhibited) when the librator tilts to the right, and tunneling on the left is promoted (and tunneling on the right inhibited) when the librator tilts to the left.
Symmetric coupling, on the other hand, describes the circumstance in which tunneling between a molecule and its neighbors on both the left and right is promoted during the same phase of oscillation and inhibited during the complementary phase; this may happen, for example, if the strength with which a mobile exciton is bound varies with the coordinate described by the oscillator.
In this paper, we restrict our attention to antisymmetric nonlocal coupling.

In momentum space, these several terms take the form
\begin{equation}
\hat{H}^{ex} = \sum_k J_k a_k^{\dagger}  a_k
\end{equation}
\begin{equation}
\hat{H}^{ph} = \sum_q \omega b_q^{\dagger} b_q
\end{equation}
\begin{equation}
\hat{H}^{ex-ph} = N^{-1/2} \sum_{k q} \omega f _{- k}^q a_{k+q}^{\dagger} a_k ( b_q + b_{- q}^{\dagger} )
\end{equation}
where
\begin{equation}
J_k = - 2 J \cos k , ~~~ 
f^q_k = g+i\phi[\sin k-\sin(k-q)],
\end{equation}

Throughout this paper, we use the the Fourier conventions for ladder operators $c^{\dagger} = a^{\dagger} , b^{\dagger}$, etc. and scalars $\gamma = \alpha , \beta $, etc.
\begin{equation}
c^{\dagger}_n  = N^{-1/2} \sum_p e^{-ipn} c^{\dagger}_p, ~~~
c^{\dagger}_p  = N^{-1/2} \sum_n e^{ipn} c^{\dagger}_n,
\end{equation}
\begin{equation}
\gamma_n ~=~ N^{-1} \sum_p e^{ipn} \gamma_p, ~~~~
\gamma_p ~=~ \sum_n e^{-ipn} \gamma_n,
\end{equation}
To assist in the interpretation of the various formulas to follow, we denote exciton wave vectors by latin $k$'s, phonon wave vectors by latin $q$'s, and reserve the greek $\kappa$ for the total crystal momentum label.

Also throughout this paper, we use the Einstein frequency as the scale of the energy, so that all energies are dimensionless.
This leaves us with a three-dimensional parameter space in which a general point may be identified as $( J , g , \phi )$.

A complete exploration of the $(J, g, \phi)$ parameter space is far beyond the scope of this paper, so we must narrow our focus significantly.
Since our principal interest here is in nonlocal exciton-phonon coupling, we defer an in-depth discussion of the $( J, g , 0 )$ plane to other work.
Since we wish to compare our present results with those of the Munn-Silbey method, we present results for the $( 0, g, \phi )$ plane in a manner that closely follows prior work.
Since the regime of strong nonlocal coupling has received much less attention than that of local coupling, we present a more systematic survey of the $(J, 0, \phi )$ plane, including a phase diagram and a discussion of characteristic polaron structures and the nonlocal-coupling version of the self-trapping transition.
Finally, we present sample results for general points in the $(J, g, \phi )$ space.

The layout of the paper is as follows.
As a preliminary study, we briefly touch upon the application of the small polaron Ansatz to nonlocal exciton-phonon coupling, as was originally suggested by Merrifield in his 1964 paper \cite{Merrifield64}, but was never carried out.
Toyozawa's Ansatz is then introduced to examine the same problem in greater depth.
The rest of the paper is devoted to exploring the parameter space under Toyozawa's Ansatz, and a comparison of our present results with comparable results of the Munn-Silbey approach \cite{Zhao94a} is made.
We must emphasize that these comparisons are not to the original analytical calculations of Munn and Silbey \cite{Munn85}, but to our own implementation of their approach by numerical methods \cite{Zhao94a}.
Since the results of our numerical study differed quantitatively and qualitatively from the original calculations of Munn and Silbey in some significant respects and thereby generalized them, it is important in the following to distinguish between the Munn-Silbey {\it method} and our own numerical {\it results} obtained by this method.

\section{The Small Polaron Ansatz} 

As in Merrifield's original calculation, the normalized small polaron trial state may be written
\begin{equation}
| \Psi (\kappa) \rangle  = N^{-1/2} \sum_n e^{i \kappa n} a^{\dagger}_n \exp [ - \sum_{n_2} ( \beta_{n_2 -n}^\kappa b^{\dagger}_{n_2} - \beta_{n_2 -n}^{\kappa \ast} b_{n_2 } )] |0\rangle, 
\end{equation}
\begin{equation}
| \Psi ( \kappa ) \rangle  = N^{-1/2} \sum_n e^{i \kappa n} a^{\dagger}_n \exp [ - N^{-1/2} \sum_q ( \beta^\kappa_q e^{-iqn}
 b^{\dagger}_q - \beta^{\kappa \ast}_q e^{iqn} b_q )] |0\rangle .
\end{equation}
This trial state may be viewed as a phased sum over ``form factors'' describing local exciton-phonon correlations.
The form factor in this case is the product of an exciton completely localized on a single lattice site and a lattice function defined {\it relative to} the exciton.

Using this trial state, we form the expectation value of the Hamiltonian and minimize the total energy $ E^\kappa $ with respect to the phonon amplitudes, yielding the self-consistency equations
\begin{equation}
\label{eq:sce}
\beta^\kappa_q  = \frac { g + i 2 \phi S^\kappa \cos(\kappa - \Phi^\kappa - \frac q 2 ) \sin  \frac q 2 }
{ 1 - 4 J S^\kappa \sin (\kappa- \Phi^\kappa  - \frac q 2 ) \sin \frac q 2 
- 8 \phi S^\kappa R^\kappa_q \sin \frac q 2 } ,
\end{equation}
\begin{equation}
S^\kappa  =  \exp [ N^{-1} \sum_q | \beta_q^\kappa |^2 (\cos q -1) ] ,
\end{equation}
\begin{equation}
\Phi^\kappa  = N^{-1} \sum_q | \beta_q^\kappa |^2 \sin q ,
\label{eq:phi}
\end{equation}
\begin{equation}
R^\kappa_q    = N^{-1} \sum_{q  ^\prime}
 Im ( \beta_{q ^\prime}^\kappa ) \sin ( \kappa - \Phi^\kappa - \frac q 2
- \frac {q ^\prime} 2 ) \sin \frac {q ^\prime} 2 .
\end{equation}
in which $S^\kappa$ and $\Phi^\kappa$ are the magnitude and phase of the Debye-Waller factor (see Section III).

It is evident from (\ref{eq:sce}) that unlike the original work of Merrifield restricted to local coupling only \cite{Merrifield64}, $\beta ^\kappa_q$ here can not be taken to be real owing to the presence of nonlocal exciton-phonon coupling.
The phonon mode amplitude $\beta^\kappa_q$ is real for all $\kappa$ and $q$ when $\phi = 0$; when $\phi \not =0$, $\beta_q^{\kappa}$ is real only along the lines defined by $ [ \sin ( \kappa - \Phi^\kappa ) - \sin ( \kappa -\Phi^\kappa - q ) ] = 0 $, among which is the $ q = 0 $ line.

Using a numerical iteration scheme, we may solve the above set of self-consistency equations to desired precision without any artificial constraints.
For all values of local and nonlocal coupling strengths, we found our final solutions to be insensitive to our choice of both numerical method and states used to initialize calculation.
For the purpose of illustration, we may express $ \beta^\kappa_q $ in terms of two real matrices $\xi^ \kappa_q, ~\eta^\kappa_q $ as in Ref~\cite{Zhao94a}:
\begin{equation}
\beta^\kappa_q  = g \xi^\kappa_q + i  \phi \eta^\kappa_q
[ \sin ( \kappa - \Phi^\kappa ) - \sin ( \kappa -\Phi^\kappa - q ) ].
\end{equation} 
From Eq.~(\ref{eq:sce}) it follows that $\eta^\kappa_q  = \xi^\kappa_q S^\kappa$.
The $q$-independence of $S^\kappa$ suggests that the $\xi$ and $\eta$ matrices may be more similar in shape than we found in our numerical implementation of the Munn-Silbey method \cite{Zhao94a}; moreover, since $\kappa$-dependence of $S^\kappa$ can be relatively weak in some regimes, the difference between $\xi$ and $\eta$ can be quite small.

\begin{figure}[htb]
\begin{center}
\leavevmode
\epsfxsize = 4.8in
\epsffile{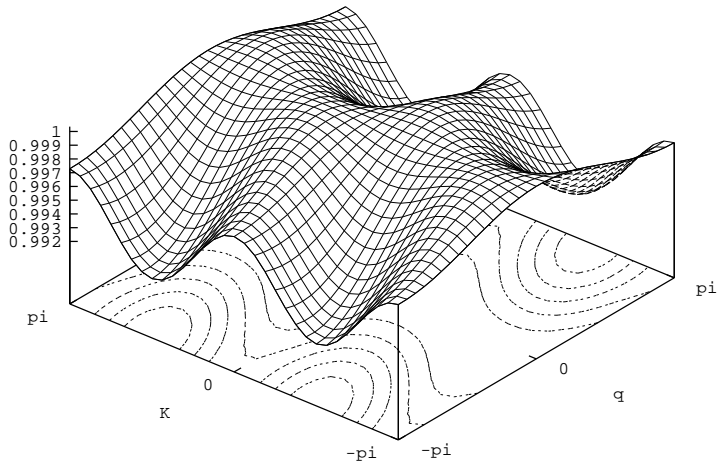}
\end{center}
\caption{
The phonon displacement factor $\xi^\kappa_q$ calculated from the small polaron Ansatz for $( J, g, \phi ) = ( 0, 1, 0.1 )$.
The variation in $\xi_q^\kappa$ is small in this case because the nonlocal coupling strength $\phi$ is weak relative to the local coupling strength $g$.
}
\label{f4.1}
\end{figure}

Results of such calculations for weak nonlocal coupling $( 0, 1, 0.1)$ are indicated in Fig.~\ref{f4.1}.
Only $ \xi^\kappa_q $ is displayed since $S^\kappa$ in this case varies by only a fraction of a percent across the entire Brillouin zone so that $\eta_q^\kappa$ would be indistinguishable from $\xi_q^\kappa$.
Since nonlocal coupling is weak relative to local coupling in this particular example, it follows as well that $\beta_q^\kappa ~ \approx ~ \xi_q^\kappa$.
The weak asymmetry of $\beta_q^\kappa$ ($\xi_q^\kappa$) with respect to $q$ at intermediate values of $\kappa$ implies (through Eq. \ref{eq:phi}) that the Debye-Waller phase $\Phi^\kappa$ varies slightly with $\kappa$; however, this weak non-constancy of $\Phi^\kappa$ has little direct influence over the properties of the solution both because of the smallness of $\Phi^\kappa$ and the small value of the nonlocal coupling constant.

Antisymmetric nonlocal exciton-phonon coupling introduces characteristic distortions in quantities such as $\xi^\kappa_q$ and the polaron energy band.
These distortions are ``bimodal'' with respect to the total crystal momentum $ \kappa $ in the sense that this modulation characteristically is most pronounced at intermediate values of $| \kappa |$.
With respect to the phonon wave vector, these distortions characteristically are most pronounced at $ |q|=\pi$, indicating that the strongest exciton-phonon coupling involves lattice dimerization.
On the other hand, $\xi^\kappa_q$ is weakly structured at low phonon wave vectors, and is in fact equal to unity along $q=0$, consistent with the lattice ``sum rule''
\begin{equation}
\label{eq:sumrule}
\beta_{q = 0} = \sum_n \beta_n = g ~ .
\end{equation}

\begin{figure}[htb]
\begin{center}
\leavevmode
\epsfxsize = 3.2in
\epsffile{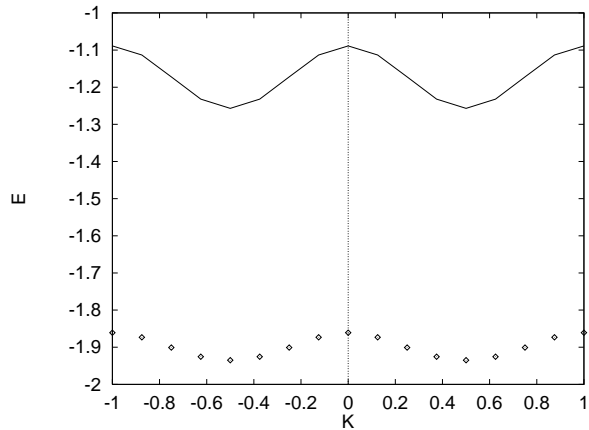}
\end{center}
\caption{
Comparison of the polaron bands between the Munn-Silbey approach \protect\cite{Zhao94a}\protect (points) and the small polaron Ansatz (solid line) for $( J, g, \phi ) = ( 0, 1, 1,)$ at $T=0$.
The small polaron Ansatz gives a higher ground state energy and a larger bimodal variation in the polaron band.
}
\label{f4.2}
\end{figure}

The polaron band energy $ E^\kappa $ can be calculated from
\begin{eqnarray}
E^\kappa & = &
N^{-1} \sum_q  | \beta_q |^2 - 2J 
S^\kappa \cos(k- \Phi^\kappa ) - 2N^{-1} g \sum_q  Re( \beta^\kappa_q ) \nonumber \\
& & - 4 N^{-1} \phi S^\kappa  \sum_q   Im( \beta^\kappa_q )
\cos ( \kappa - \Phi^\kappa - \frac q 2 ) \sin ( \frac q 2 ) .
\end{eqnarray}
Comparison between the small polaron band and that calculated numerically from the Munn-Silbey approach \cite{Zhao94a} is shown in Fig. \ref{f4.2} for the case of $( J, g, \phi ) = ( 0, 1, 1 )$.

The Munn-Silbey method yields a flatter and lower energy band than does the small-polaron method; however, since the perturbative result falls below the variational result, no conclusion can be drawn as to which approach gives the better polaron band; for example, without further information we cannot exclude the possibility that the actual ground state band might lie between these two results and closer to the variational band.
Higher-quality variations are needed to make such determinations.

Through its Fourier transform, the nonuniformity of $ \xi^\kappa_q $ with respect to $q$ (though small on an absolute scale in this case) implies a spreading of phonon distortion in site space even in the absence of a direct transfer integral $J$.
With increasing nonlocal coupling, the amplitude of this distortion increases, as does its spread in real space.

Since nonlocal exciton-phonon coupling provides the only transport mechanism when $J=0$, it is reasonable to expect in analogy with the usual small polaron picture (e.g., as in Ref.~\cite{Merrifield64}) that a small polaron approach such as this one may break down when nonlocal coupling becomes comparable in strength to local coupling.

At the least, the approach sketched here needs to be generalized to admit wave function symmetries more consistent with the dimeric structures favored by nonlocal coupling.
This could be done in a minimal way by building up trial Bloch states from dimeric form factors; we elect, however, to skip this step and proceed to a still more flexible trial state introduced by Toyozawa that is well known in the context of local-coupling polaron theory. 

\section{Toyozawa's Ansatz}

The major shortcoming of the small polaron Ansatz is the fact that it is built up from form factors in which the electronic component is completely localized on a single lattice site.
In the presence of significant exciton tunneling, or, as we have observed, in the presence of significant nonlocal exciton-phonon coupling, it would appear reasonable to expect self-consistent local exciton-phonon correlations to reflect at least some of the characteristics of spreading wave packets.
To accommodate this expecatation, the completely-localized exciton component of the small polaron form factor can be generalized to a superposition of exciton amplitudes to be determined self-consistently with the amplitudes of the lattice oscillators.
This is the basis of many theories characterizable as ``large polaron'' theories.
Important among these for our purposes are delocalized-state theories based on phased sums of such generalized form factors (historically identified with Toyozawa), and localized-state theories based upon using such generalized form factors as trial states in their own right without invoking the phased-sum construction that assures delocalization (historically identified with Pekar
\cite{Pekar46a,Pekar46b,Landau48,Pekar54}
and Davydov
\cite{Davydov69,Davydov73,Davydov76,Davydov79,Fischer84,Venzl85,Skrinjar88,Zhang88,Brown89,Ivic89,Wang89,Brown90,Wang90}).

Toyozawa's Ansatz may be viewed as a delocalization of the Davydov Ansatz $| \psi \rangle$
\begin{equation}
| \psi \rangle ~=~ | \alpha \rangle \otimes | \beta \rangle ,
\end{equation}
where $\otimes$ denotes the direct product, and $| \alpha \rangle$ and $| \beta \rangle$ are the exciton and phonon part of the form factor respectively, built from exciton and phonon vacua $|0\rangle_{ex}$ and $|0\rangle_{ph}$ as
\begin{equation}
| \alpha \rangle = \sum_n  \alpha_n a_n^{\dagger} |0\rangle_{ex} ,
\end{equation}
\begin{equation}
| \beta \rangle = \exp [ { \sum_n ( \beta_n b_n^{\dagger} - 
\beta_n^\ast b_n )} ] |0\rangle_{ph} .
\end{equation}
After delocalization one obtains Toyozawa's Ansatz state, given by
\begin{equation}
| \Psi ( \kappa ) \rangle = | \kappa \rangle \langle \kappa | \kappa \rangle^{-1/2}
\end{equation}
\begin{equation}
| \kappa \rangle = \sum_{n n_1} e^{i \kappa n} \alpha_{n_1 - n}^\kappa
a_{n_1}^{\dagger}
\exp [ - \sum_{ n_2} (\beta_{n_2 -n}^\kappa
b_{n_2}^{\dagger} -\beta_{n_2 -n}^{ \kappa \ast} b_{n_2 } )] |0\rangle
\end{equation}
\begin{equation}
| \kappa \rangle = N^{-1/2} \sum_{nk} e^{i( \kappa -k)n} \alpha_k^\kappa a_k^{\dagger}
\exp [ - N^{-1/2} \sum_q ( \beta_q^\kappa e^{-iqn}
 b_q^{\dagger} -\beta_q^{ \kappa \ast} e^{iqn} b_q )] |0 \rangle .
\end{equation}
where $|0\rangle = | 0 \rangle_{ex} \otimes | 0 \rangle_{ph}$.
The auxiliary vector $| \kappa \rangle$ is not normalized, but simplifies presentation of some results.
It may be worth noting that many localized-state approaches such as the Davydov theory take the exciton component $| \alpha \rangle$ to be normalized such that $\sum_n | \alpha_n |^2 = 1$, since in this approach the norm of exciton amplitudes has a physical interpetation as the exciton number.
In the delocalized-state approach, however, the norm of the exciton amplitudes does not have this physical meaning, and by contributing only a multiplicative factor to $\langle \kappa | \kappa \rangle$, makes no contribution to the normalized trial state.
Consequently, the norm of the amplitudes $\alpha$ is arbitrary, and in fact drifts in the course of numerical variation; since this drift has no physical consequences, we have found it convenient throughout this paper to normalize the exciton amplitudes such that $\alpha_{n=0} = 1$.

For variational calculations we require the total energy associated with each trial state, for which we evaluate the expectation values of the three principal terms in the Holstein Hamiltonian:
\begin{equation}
\langle \kappa | \hat{H}^{ex} | \kappa \rangle = -2 J N^{-1} \sum_k S_{ \kappa - k
}^\kappa \cos k | \alpha_k^\kappa |^2 ,
\end{equation}
\begin{equation}
\langle \kappa | \hat{H}^{ph} | \kappa \rangle = N^{-2} \sum_{kq}
S_{ \kappa - k - q }^\kappa | \alpha_k^\kappa |^2
|\beta_q^\kappa |^2 ,
\end{equation}
\begin{equation}
\langle \kappa | \hat{H}^{ex-ph} | \kappa \rangle = - N^{-2} \sum_{kq}
f^{-q}_{-k-q}
\alpha_k^{\kappa \ast} \alpha_{k+q}^\kappa
( S_{ \kappa - k - q }^\kappa\beta_q^{\kappa \ast} + S_{ \kappa - k }^\kappa\beta_{-q}^\kappa ) .
\end{equation}
\begin{equation}
\langle \kappa | \kappa \rangle  = N^{-1} \sum_k
S_{ \kappa - k }^\kappa | \alpha_k^\kappa |^2 .
\end{equation}
Here, $S_k^\kappa$ is the Fourier transform of the generalized Debye-Waller factors $S_n^\kappa$
\begin{equation}
\label{eq:skk}
S_k^\kappa= \sum_n e^{-ikn} S_n^\kappa,
\end{equation}
\begin{equation}
\label{eq:skn}
S_n^\kappa =  \exp [ N^{-1} \sum_q |\beta_q^\kappa |^2 (e^{iqn} -1) ],
\end{equation}
which quantifies the overlap between the lattice components of polaron wavefunctions displaced from each other by $n$ lattice sites.
These are to be distinguished from the Franck-Condon factor
\begin{equation}
_{ph}\langle 0 | \beta \rangle = e^{- \frac 1 2 \sum_n |\beta_n |^2}
,
\end{equation}
which quantifies the overlap between the lattice component of the polaron wavefunction with the undistorted phonon ground state.
The Debye Waller factors $S_{n = \pm 1}^\kappa$ ($= S^{\kappa} e^{\pm i \Phi^\kappa}$ of section II), appear routinely in the transport terms of effective (small) polaron Hamiltonians, where they strongly influence the renormalization of the effective mass.
In the general case we address here, the spread of the exciton amplitudes $\alpha_n$ causes Debye-Waller factors between non-nearest-neighbor sites to contribute to polaron structure in complex ways.

Owing to the large number of Debye-Waller factors that may be involved in the general case, their resolution into magnitudes and phases as in section II ceases to be advantageous.
One can show from (\ref{eq:skk}) and (\ref{eq:skn}) that $S^{\kappa}_{k}$ is strictly real, so that computation is simplified by abandoning the magnitude and phase characterization and working in momentum space with the real quantities $S_k^\kappa$.

Minimization of the total energy $E^\kappa$ with respect to the phonon amplitudes $\beta_q^{ \kappa \ast}$ leads to
\begin{equation}
\label{eq:beta}
\beta_q^\kappa ~=~ \frac { L_q^\kappa}
{ M_q^\kappa  ~+~ H_q^{\kappa} ~-~ M_q^\kappa E^\kappa }
\end{equation}
where
\begin{equation}
L_q^\kappa = N^{-1} \sum_k f^{-q}_{-k-q} S_{ \kappa - k - q }^\kappa
\alpha_k^{ \kappa \ast} \alpha_{k+q}^\kappa,
\end{equation}
\begin{equation}
M_q^\kappa  = N^{-1} \sum_k S_{ \kappa - k - q }^\kappa
| \alpha_k^\kappa |^2,
\end{equation}
and $H_q^\kappa$ is the sum of three terms:
\begin{equation}
H_q^{\kappa} = H_q^{ex} + H_q^{ph} + H_q^{ex-ph},
\end{equation}
\begin{equation}
H_q^{ex}  = -2J N^{-1} \sum_k
S_{ \kappa - k - q }^\kappa
\cos k | \alpha_k^\kappa |^2,
\end{equation}
\begin{equation}
H_q^{ph} = N^{-2} \sum_{kq^\prime}
S_{ \kappa - k - q - q^\prime }^\kappa
| \alpha_k^\kappa |^2  |\beta_{q^\prime} ^\kappa |^2,
\end{equation}
\begin{equation}
H_q^{ex-ph} = - N^{-2} \sum_{kq^\prime}
f^{-q^\prime}_{-k-q^\prime}
\alpha_k^{ \kappa \ast} \alpha_{k+q^\prime}^\kappa
( S_{ \kappa - k - q - q^\prime }^\kappa\beta_{q^\prime}^{ \kappa \ast}
+ S_{ \kappa - k - q }^\kappa\beta_{-q^\prime}^\kappa ).
\end{equation}

Similarly, after minimizing $E^\kappa$ with respect to $ \alpha_k^{ \kappa \ast }$ we arrive at
\begin{equation}
\alpha_k^\kappa = { { L_k^\kappa } \over
{ M_k^\kappa - ( E^\kappa + 2J \cos k ) S_{ \kappa - k }^\kappa }}
\end{equation}
where
\begin{equation}
L_k^\kappa =  N^{-1} \sum_q f^{-q}_{-k-q} \alpha_{k+q}^\kappa
( S_{ \kappa - k - q }^\kappa\beta_q^{ \kappa \ast} + S_{\kappa-k}^\kappa\beta_{-q}^\kappa )
\end{equation}
and
\begin{equation}
M_k^\kappa = N^{-1} \sum_q S_{ \kappa - k - q }^\kappa
|\beta_q^\kappa |^2 .
\end{equation}

We may recover the lattice sum rule (\ref{eq:sumrule}) from (\ref{eq:beta}) by noting that
\begin{equation}
H_{q=0}^{\kappa} = E^\kappa \langle \kappa | \kappa \rangle ~,~~~~~
g^{-1} L_{q=0}^\kappa = M^\kappa_{q=0} = \langle \kappa | \kappa \rangle .
\end{equation}

\section{\lowercase{$g$} and $\phi$}

The effect of nonlocal exciton-phonon coupling is most evident when $J$ is small or absent.
Moreover, since local and nonlocal coupling influence polaron structure in distinct ways, it is well to examine the influence of nonlocal coupling both in isolation and in concert with local coupling.

\begin{figure}[htb]
\begin{center}
\leavevmode
\epsfxsize = 3.2in
\epsffile{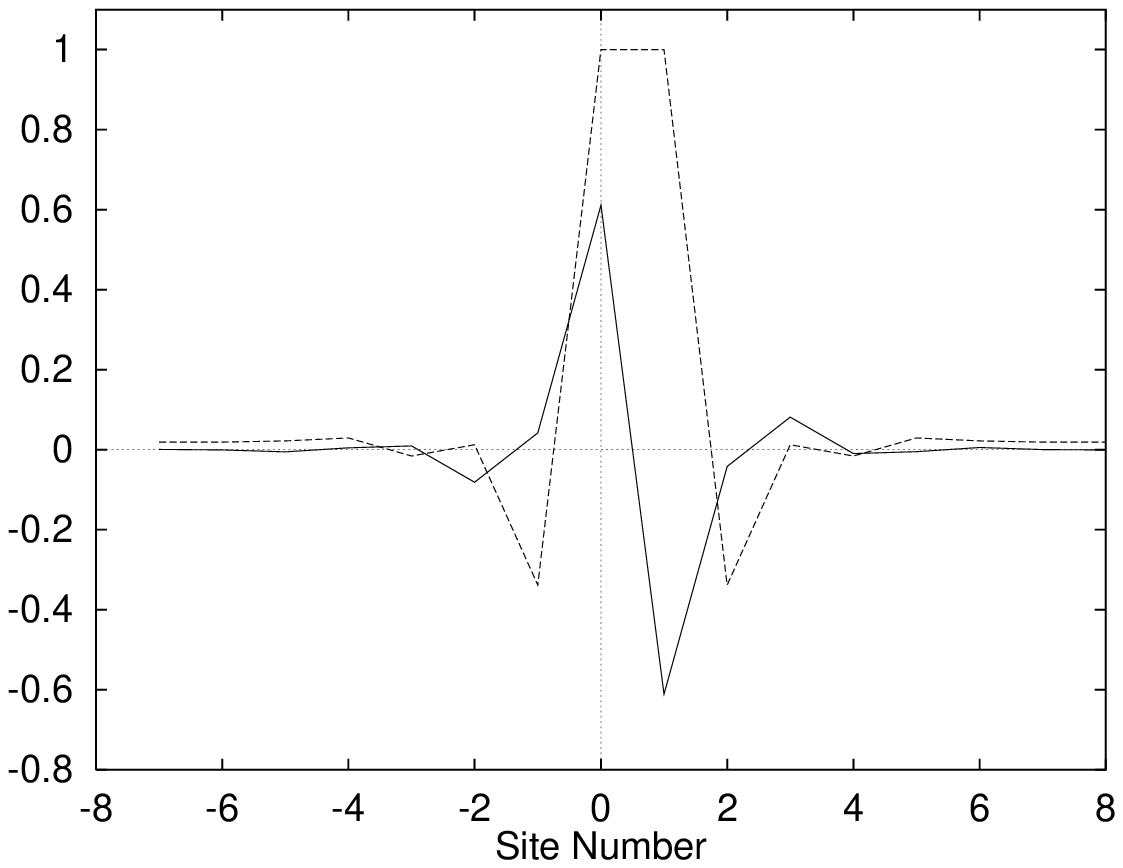}
\vspace{.1in}
\leavevmode
\epsfxsize = 3.2in
\epsffile{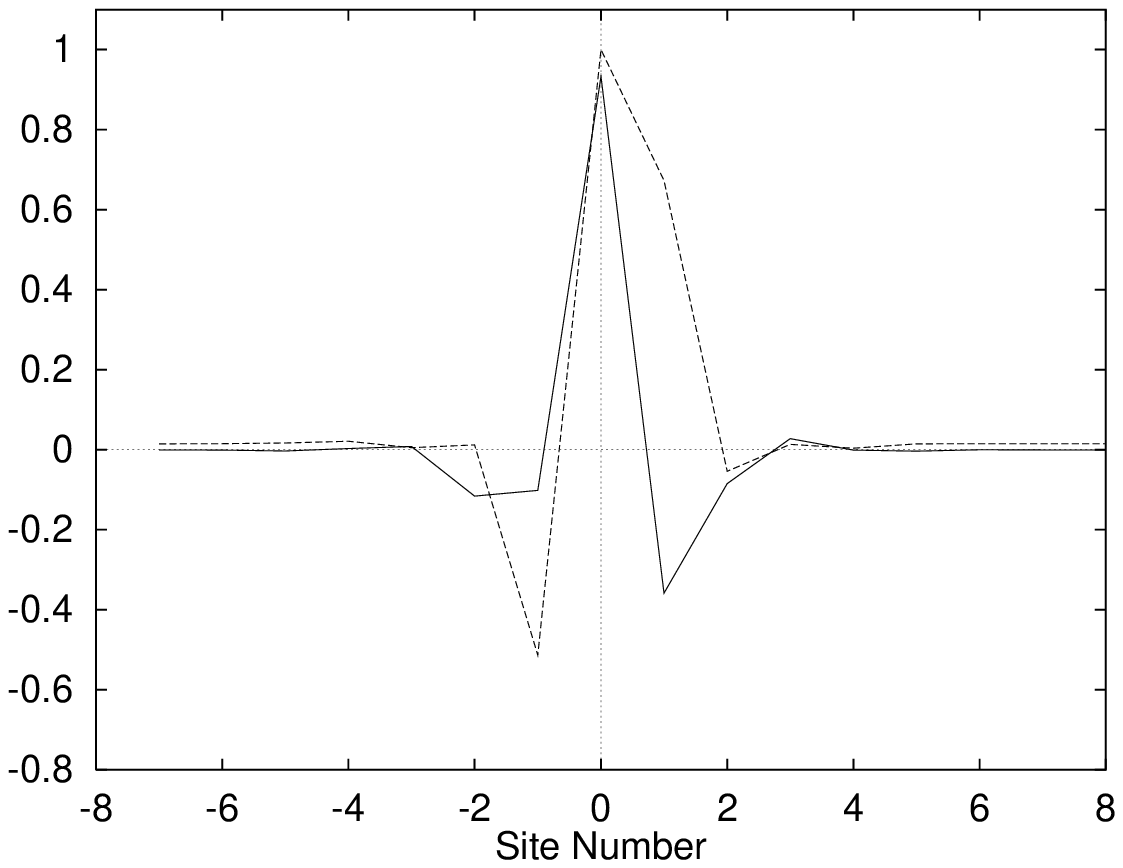}
\end{center}
\caption{
Variational parameters for the $\kappa~=~0$ state calculated from Toyozawa's Ansatz.
Second panel: the exciton amplitude $\alpha_n^{\kappa=0}$ (dashed line) and the phonon displacement $\beta_n^{\kappa=0}$ (solid line) for $( J, g, \phi ) = ( 0, 0, 1 )$;
First panel: the exciton amplitude $\alpha_n^{\kappa=0}$ (dashed line) and the phonon displacement $\beta_n^{\kappa=0}$ (solid line) for $( J, g. \phi ) = ( 0, 0.3, 1 )$.
}
\label{f4.3}
\end{figure}

The first panel of Fig.~\ref{f4.3} shows the exciton and phonon amplitudes $\beta^{\kappa}_n$ and $\alpha^{\kappa}_n$ for the pure nonlocal coupling case $( J , g , \phi ) = ( 0 , 0 , 1 )$ at $\kappa = 0$ where both of these quantities are real.

In the absence of local coupling, the exciton amplitudes $\alpha^{\kappa=0}_n$ are bond-centered and even under inversion through the occupied bond, while $\beta^{\kappa=0}_n$ is bond-centered and odd under inversion, reflecting the antisymmetry of the nonlocal exciton-phonon coupling.
Approximately 84\% of the total exciton density is shared equally by the two sites defining the central bond.
This is characteristic of the strong nonlocal coupling regime, and justifies some the assumptions of our previous work.

The first panel of Fig.~\ref{f4.3} should be compared to the exciton-phonon correlation function $\bar{C}_l$ produced as a diagnostic of polaron structure in our previous paper \cite{Zhao94a};
$\bar{C}_l$ was constructed around a dimeric exciton function $\tilde{\psi}_n$ restricted to two sites (i.e., 100\% of the total exciton density shared equally between $n=0,1$) that we presumed to be representative of the exciton amplitudes $\alpha^{\kappa=0}_n$ in actual polaron states.
The similarity between $\beta^{\kappa=0}_n$ of Fig.~\ref{f4.3}a and $\bar{C}_l$ is striking.
(The previous calculation included a small local coupling strength that does not significantly affect the comparison.)

The second panel of Fig.~\ref{f4.3} shows $\beta^{\kappa=0}_n$ and $\alpha^{\kappa=0}_n$ for $( J, g, \phi ) = ( 0, 0.3, 1.0 )$, reflecting the addition of a moderate amount of local coupling to the previous scenario.
It is still the case that most of the total exciton density resides on the two sites defining the central bond (this does not change materially until $J$ becomes signficant); however, the addition of a non-negligible amount of local coupling destroys the bond-centered symmetry by introducing a site-centered structure localized on only one of the two central sites.

\begin{figure}[htb]
\begin{center}
\leavevmode
\epsfxsize = 3.2in
\epsffile{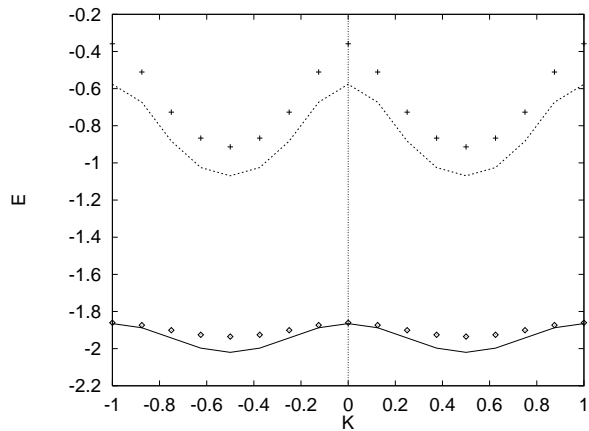}
\end{center}
\caption{
Comparison between the polaron band calculated from the Munn-Silbey approach \protect\cite{Zhao94a}\protect at $T=0$, (diamonds, $( 0, 1, 1 )$, crosses, $( 0, 0.03, 1)$ and that from Toyozawa's Ansatz (solid line, $( 0, 1, 1 )$, dashed line, $( 0, 0.03, 1)$.
}
\label{f4.8}
\end{figure}

Fig.~\ref{f4.8} displays the energy band comparison between the Munn-Silbey approach \cite{Zhao94a} and Toyozawa's Ansatz for $( 0, 0.03, 1)$ (essentially Figure~\ref{f4.3}a) and $(0,1,1)$, both of which were studied in Ref.~\cite{Zhao94a}.

Though still imperfect, the energy band comparison is much more favorable here than we found in section II.
The consonance between our present results and our previous calculations by the Munn-Silbey method appears to be due to the ability of both approaches to embrace dimeric exciton-phonon correlations essential to nonlocal-coupling polarons.
Our present results based on Toyozawa's Ansatz give the best estimate of the polaron ground state energy among the three we consider, and can be viewed as corroborating our prior numerical calculations by the Munn-Silbey method, at least when nonlocal coupling dominates.
Due to the variational nature of Toyozawa's Ansatz, it is now safe to say that the Munn-Silbey approach over estimates the ground state energy at $T=0$.

\section{$J$ and $\phi$}

In this section we put local coupling aside (i.e., we set $g=0$), and focus on the interplay between exciton tunneling ($J$) and nonlocal coupling ($\phi$).
We have seen that although nonlocal coupling is an exciton-phonon interaction that supports polaron formation, it is also a transport mechanism and competes with local coupling both by promoting transport and by driving the exciton-lattice correlations toward dimeric structures rather than the site-localized structures preferred by local coupling interactions.
On the other hand, although nonlocal coupling is a transport mechanism, it also competes with direct, phonon-free exciton transfers since the lattice distortions inherent in phonon-assisted transfers inhibit direct transfers.

\begin{figure}[htb]
\begin{center}
\leavevmode
\epsfxsize = 3.2in
\epsffile{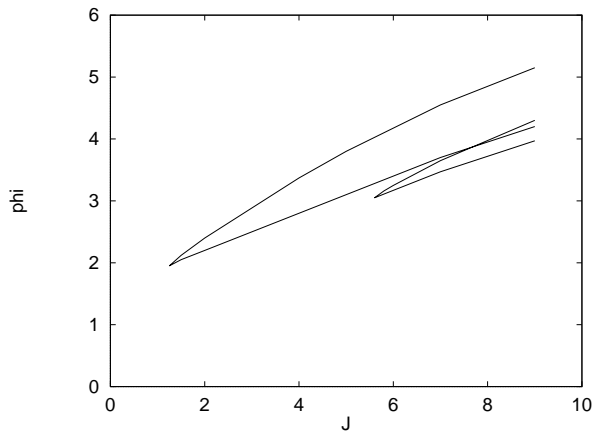}
\end{center}
\caption{
Phase diagram on the $J - \phi$ plane for Toyozawa's Ansatz.
The two wedges correspond to the discontinuity near the Brillouin zone center (upper wedge) and that near the Brillouin zone boundary (lower wedge), respectively.
}
\label{f4.5}
\end{figure}

It is convenient to use the device of a phase diagram to organize our discussion.
Fig~\ref{f4.5} depicts the $J - \phi$ plane at $g=0$, in which we have indicated two narrow, tongue-shaped regions that divide the plane into two more-or-less distinct regions, a {\it strong} nonlocal coupling region occupying the upper portion of the diagram and a {\it weak} nonlocal coupling region occupying the lower.
The phenomena indicated by these tongues can be characterized as a kind of self-trapping.

In the more familiar self-trapping phenomenon associated with local exciton-phonon coupling, a polaron is understood to be characterized by exciton-phonon correlations that are more-or-less spatially compact.
For ``weak'' local coupling, correlations may extend over many lattice sites, corresponding to a ``large'' polaron.
With increasing local coupling strength, the width of the correlated region diminishes until the excitation is essentially confined to a single lattice site, corresponding to a ``small'' polaron.
Along the way, other polaron properties change as well, most notably the polaron effective mass growing from the free mass at weak coupling to values arbitrarily large in the strong coupling region.
Whether this change as a function of system parameters is necessarily smooth or may occur discontinuously is a subject of some contention \cite{Gerlach87i,Gerlach87ii,Lowen88,Gerlach91}; nonetheless, it is common for approximate treatments to find the change to be discontinuous in some regimes, and hence the imprecise notion of a discrete ``self-trapping transition'' is widespread.
More important than the question of continuity is the fact that in the vicinity of this transition, exciton-phonon correlations and related physical quantities such as the effective mass undergo strong changes with relatively small changes in control parameters.

The qualitative characteristics of this familiar self-trapping transition apply as well to our treatment of nonlocal coupling.
We find a discrete transition to occur in the vicinity of the Brillouin zone {\it center}, bearing some resemblance to the more familiar local coupling phenomenon; however, we also find a discrete transition occurring in the vicinity of the zone {\it edge}.
Moreover, whereas the usual conception of self-trapping focusses on dramatic changes that occur at the Brillouin zone center (e.g., the jump in the effective mass) leading to the notion of a {\it sharp} transition, we find the discrete transitions in both the inner and outer zones to be {\it broad} in a sense to be described presently.

We must now attempt to be precise in specifying the meaning of ``discrete transition'' in the present context, for clarity focussing on the inner-zone transition.

At an arbitrarily chosen point $(J,g,\phi)$ in parameter space, there exist one or more minima in the variational energy $E^\kappa$ for each $\kappa$, the lowest of which identifies the global energy minimum in each total momentum sector; the collection of energies for all $\kappa$ identifies the polaron energy band $E^\kappa$ at $(J,g,\phi)$, and the collection of {\it states} associated with these global energy minima constitute the polaron Bloch states.
The distinct classes of relative minima that may coexist over certain invervals of $\kappa$ are associated with distinct classes of variational states corresponding to different polaron structures.
In the present problem, typically, either:
1) one class of minima exists for all $\kappa$, or
2) one class of minima exists over an interval $| \kappa | \in ( 0, \kappa_1 )$, and another exists over an interval $| \kappa | \in ( \kappa_2 , \pi )$, with the two classes of relative minima coexisting for $| \kappa | \in ( \kappa_2 , \kappa_1 )$.
In the latter case, there exists a particular total crystal momentum $\kappa^* \in ( \kappa_2 , \kappa_1 )$ at which the polaron energy band changes from being defined by one class of global energy minima in the inner zone ($| \kappa | < \kappa^*$) to being defined by the other class of global energy minima in the outer zone ($| \kappa | > \kappa^*$); this $\kappa^*$, like the energy band itself, is a function of the system parameters $( J,g,\phi )$.

We describe the low-$\kappa$ states as ``large-polaron-like'' and the high-$\kappa$ states as ``small-polaron-like'' because through sequences of infinitesimal steps in $\kappa$ and/or exciton-phonon coupling parameters it is possible to smoothly deform any large-polaron-like state into a traditional large polaron state at $\kappa = 0$, and any small-polaron-like state into a traditional small polaron state at $\kappa = 0$.
Significant for the explication of the discrete transition phenomenon is the fact that it is {\it not} generally possible to smoothly deform large-polaron-like states into small-polaron-like states (or vice versa) through such sequences.
In particular, at fixed $(J,g,\phi )$ smooth changes in $\kappa$ are followed by smooth changes in all polaron properties until $\kappa$ reaches $\kappa^*$.
At $\kappa^*$, discontinuities appear in at least some polaron properities, consequent to the switching of the global energy minimum from one class of state in one part of the variational state space to another class of state finitely separated from the first.

The appearance of a $\kappa$-dependent discrete transition at $\kappa^*$ is unambiguous in our results, and by our description is a discrete transition between small- and large-polaron-like states.
What may be less clear to this point is what relationship, if any, exists between this transition phenomenon and the more traditional notion of a discrete self-trapping transition.
The ``order parameter'' commonly used as an indicator of the traditional self-trapping transition is the polaron effective mass, a quantity defined at $\kappa = 0$; as such, its characteristic jump indicates for us the point in parameter space at which the $\kappa=0$ polaron Bloch state switches from being large-polaron-like to small-polaron-like.
It is possible to connect this traditional perspective of a $\kappa=0$ transition with our present $\kappa$-dependent transition through the dependence of $\kappa^*$ upon the system parameters $(J,g,\phi)$.
The general result is that whenever a $\kappa$-dependent transition exists between small- and large-polaron-like states at a finite $\kappa^*$, there exist sequences of infinitesimal changes in the parameters $(J,g,\phi)$ that cause $\kappa^*$ to vanish.
This vanishing of $\kappa^*$ has the consequence that the large-polaron-like region ceases to exist, rendering the entire polaron band small-polaron-like, {\it including} the $\kappa = 0$ state.
This shows the usual self-trapping line to mark a {\it boundary} of the finite region of parameter space over which the $\kappa$-dependent transition exists.
This smooth connection also allows us to characterise the $\kappa$-dependent transition as a {\it continuation} to finite $\kappa$ of the traditional self-trapping transition at $\kappa~=~0$.

Where such finite-$\kappa$ transitions occur, some surgery must be performed to arrive at the polaron band that at every $\kappa$ identifies the global energy minimum.
This surgery consists of discarding the higher-lying of the coexisting variational energy minima in the interval $( \kappa_2 , \kappa_1 )$ and the states associated with them, retaining only those yielding the lowest energy at every $\kappa$.
The result is a polaron band that is large-polaron-like for $| \kappa | < \kappa^*$ and small-polaron-like for $| \kappa | > \kappa^*$.
Outside of the transition regions, polaron bands are determined without necessity of surgery; however this qualitative characterization of the $\kappa$ dependence of the polaron band as being more large-polaron-like at long wave lengths and more small-polaron-like at short wave lengths continues to apply.

The outer-zone transition is qualitatively similar in most respects, except that instead of going to zero in some limit, the $\kappa^*$ associated with the transition can be pushed to the Brillouin zone edge, and the polaron states on each side of the transition differ from those of the inner-zone transition in some characteristic ways addressed later in this section.

The tongue-shaped regions indicated on the phase diagram in Figure \ref{f4.5} identify parameters associated with such discrete transitions; the upper, wider tongue is associated with the inner-zone transition, and the lower, more narrow tongue with the outer-zone transition.
Where the tongues overlap, discrete transitions are found to occur in the inner and outer zones simultaneously, such that the same energy band embraces distinct polaron states in the inner, intermediate, and outer Brillouin zone.

The way in which these discrete transitions depend on the parameters $J$ and $\phi$ can be sketched qualitatively as follows:
For fixed $J$ sufficiently large, a finite $\kappa^*$ first appears as $\phi$ is increased from zero to the weak-coupling boundary of the zone-center transition region.
As $\phi$ is further increased, this finite $\kappa^*$ decreases until at the strong-coupling boundary of the zone-center transition $\kappa^*$ vanishes, identifying the {\it strong-coupling} boundary with the usual notion of a discrete self-trapping transition occurring at the zone center.
Transits across the outer-zone transition region have a similar structure.

We can study the nature of the ``large'' and ``small'' polaron states associated with nonlocal coupling by examining the structure of typical states at selected points in the system parameter space.

\begin{figure}[htb]
\begin{center}
\leavevmode
\epsfxsize = 3.2in
\epsffile{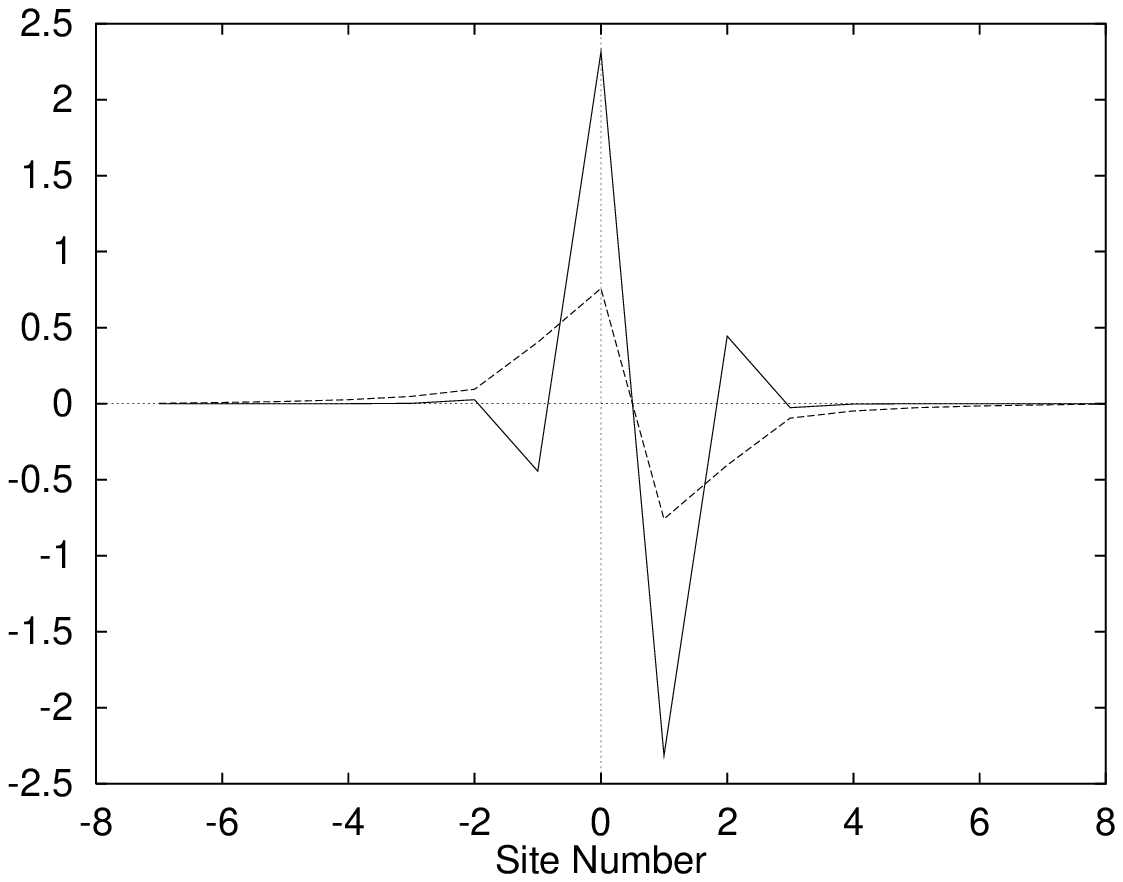}
\vspace{.1in}
\leavevmode
\epsfxsize = 3.2in
\epsffile{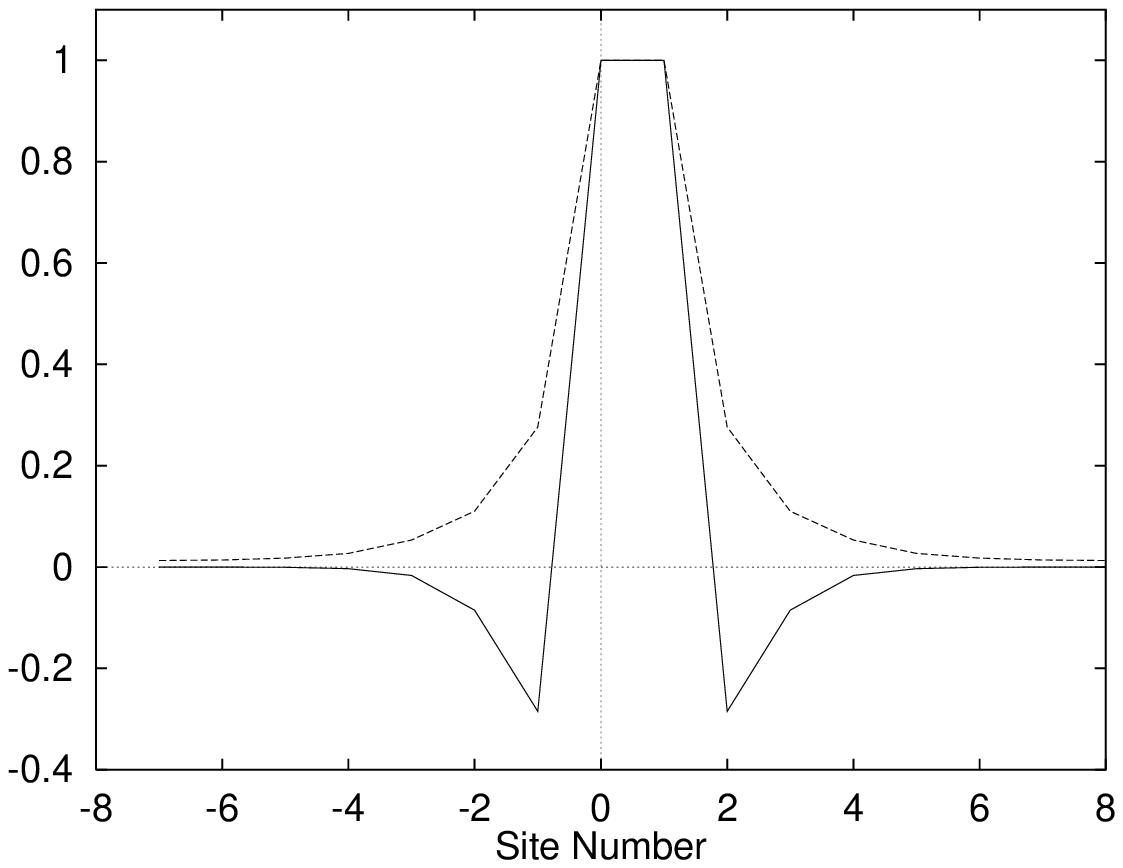}
\end{center}
\caption{
Two types of convergent solutions for $Re[ \beta^{\kappa=0}_n] $ (first panel) and $ Re[ \alpha^{\kappa=0}_n ]$ (second panel).
$( J, g, \phi ) = ( 6, 0, 4 )$.
The solid line is obtained when this point is approached from the strong coupling regime, and the dashed line is obtained when this point is approached from the weak coupling regime.
}
\label{f4.6}
\end{figure}

We turn first to the inner-zone transition and specifically the point $( J, g, \phi ) = ( 6, 0, 4 )$.
This point is near the strong-coupling edge of the transition region, where two distinct classes of stable states exist in a small interval around $\kappa = 0$.
The structure of the two stable $\kappa = 0$ solutions is illustrated in Fig.~\ref{f4.6}.

The solid lines are iterative results obtained when this point is approached from the upper portion of the phase diagram.
The dashed lines are results obtained upon approach from the lower portion of the phase diagram.
Consequently, the solid lines show the polaron structure typical of the upper portion of the $J - \phi$ plane where nonlocal exciton-phonon coupling is relatively strong, and the dashed lines show the polaron structure typical of the lower portion of the phase diagram where exciton-phonon coupling is relatively weak.
As seen in Fig.~\ref{f4.6}, the strong-coupling polaron state (solid curves) are characterized by a larger phonon displacement $\beta_n^{\kappa=0}$ and a slightly more localized exciton amplitude $\alpha_n^{\kappa=0}$ than those of the weak-coupling polaron state (dashed curves), consistent with the notion of small and large polarons, respectively.

The ``size'' of the polarons represented by the two solutions in Fig.~\ref{f4.6} do not differ markedly, although upon close observation the exciton and phonon amplitudes of the ``small polaron'' state (solid curves) do decay somewhat more rapidly than do those of the large polaron state (dashed curves).
A feature more characteristic of the difference between the large and small polaron states in the presence of nonlocal coupling is the fact that amplitudes in the small polaron state are characterized by more or stronger alternations of sign than are found in the large polaron state.
Although these alternations are ultimately due to the antisymmetric nature of the nonlocal coupling used in this paper, their greater prominance on the strong-coupling side of the zone-center transition is consistent with the more compact exciton-phonon correlations typical of small polarons.

\begin{figure}[htb]
\begin{center}
\leavevmode
\epsfxsize = 3.2in
\epsffile{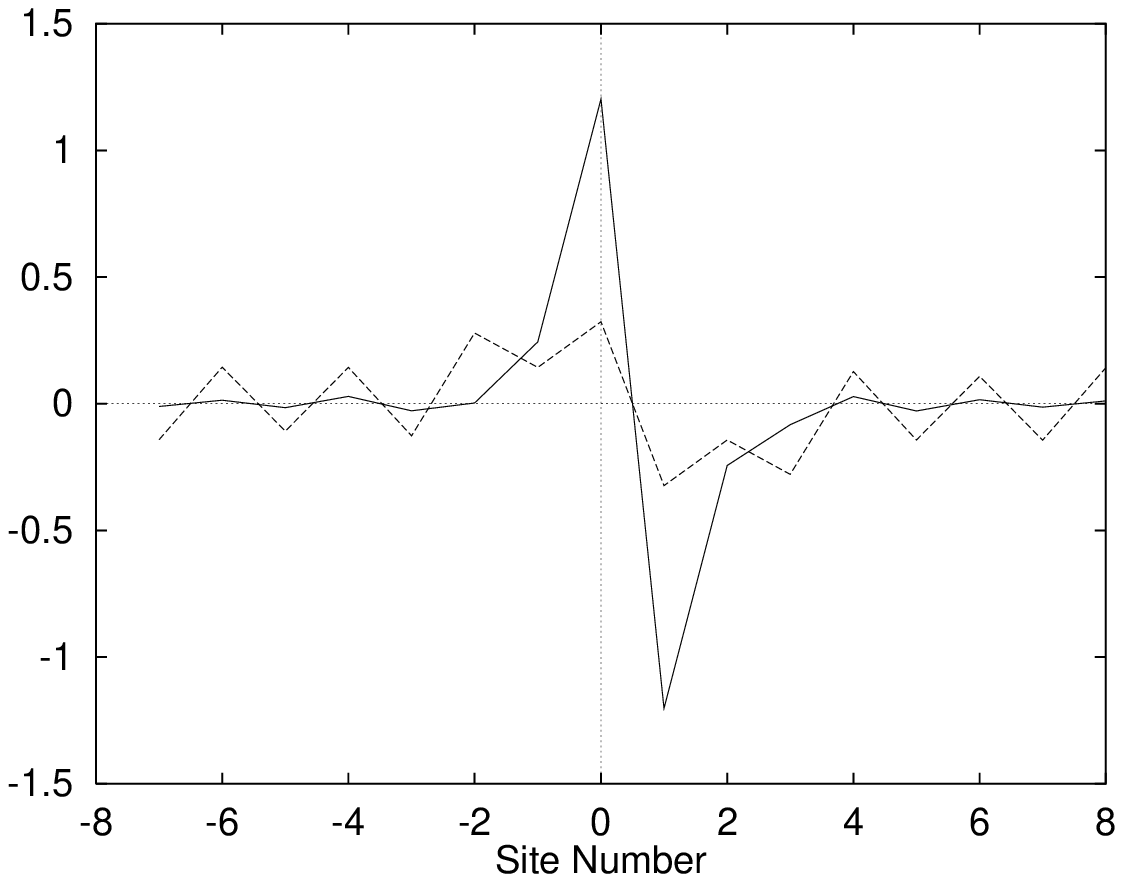}
\vspace{.1in}
\leavevmode
\epsfxsize = 3.2in
\epsffile{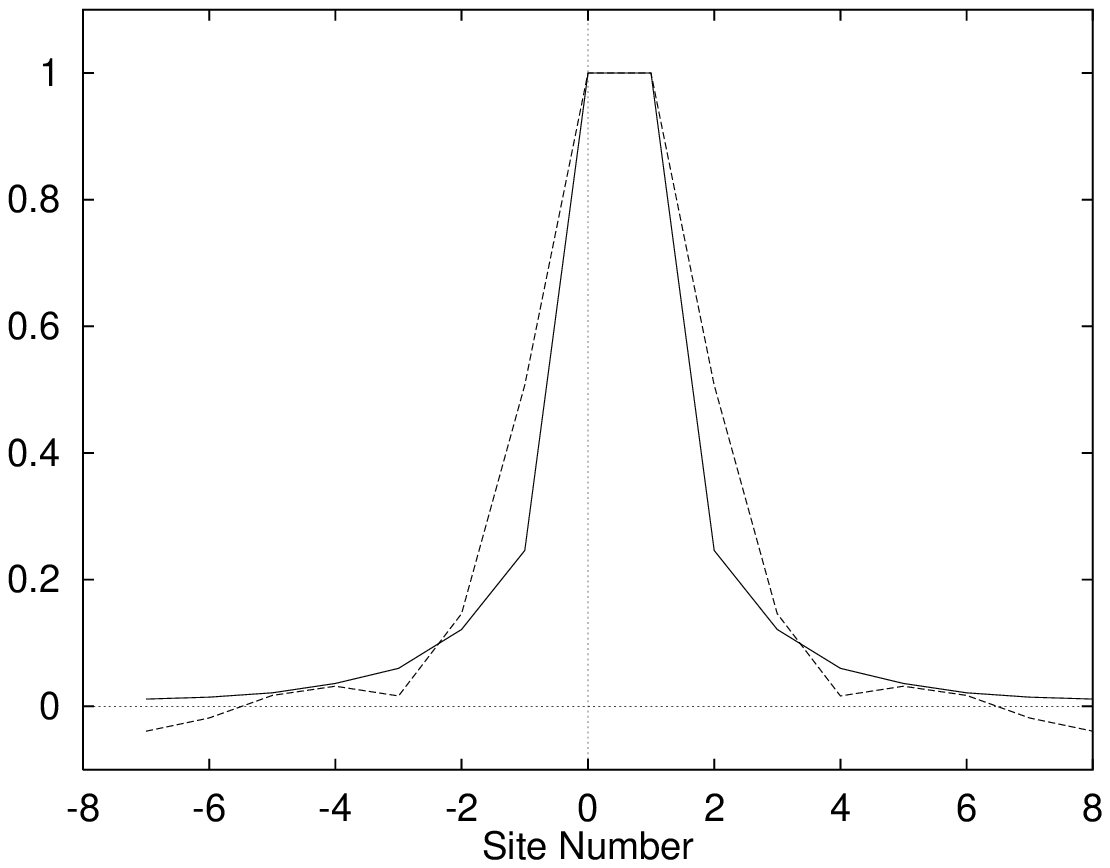}
\end{center}
\caption{
Two types of convergent solutions for $Re[ \beta ^{\kappa=\pi}_n]$ (first panel) and $ Re[ \alpha^{\kappa=\pi}_n ]$ (second panel).
$( J, g, \phi ) = ( 6, 0, 3.2 )$.
The solid line is obtained when this point is approached from the strong coupling regime, and the dashed line is obtained when this point is approached from the weak coupling regime.
In the second panel, the irregular tails of the dashed line are caused by numerical difficulties in the weak coupling regime.
}
\label{f4.10}
\end{figure}

Next we turn to the outer-zone transition, and specifically the point $( J, g, \phi ) = ( 6, 0, 3.2 )$.
This point is near the strong-coupling edge of the transition region, where two distinct classes of stable states exist in the neighborhood of $| \kappa | = \pi$.
Structures typical of these two kinds of solution are illustrated in Fig.~\ref{f4.10}.

As in Fig.~\ref{f4.6}, the solid curves in Fig.~\ref{f4.10} are obtained when this point $(J,g, \phi ) = (6,0,3.2)$ is approached from the strong-coupling regime, while the dashed curves are obtained by approaching from the weak-coupling regime.
A significant feature of the zone-edge solutions is that the phonon amplitudes can be resolved into two components; one component is a strongly localized asymmetric lattice distortion as we have seen in all the other nonlocal coupling scenarios, and the other component is a nearly-uniform plane wave.
The zone-edge transition can be viewed as a binding or unbinding of the free-phonon component suggested by this plane wave.
In the large polaron regime {\it below} the transition (higher $\kappa$, weaker coupling), the plane-wave component dominates the phonon amplitudes, while after crossing over into the small polaron regime (lower $\kappa$, stronger coupling), this plane-wave component is absent.

In the particular solution illustrated, the plane-wave component has the wave vector $q~=~\pi$, consistent with the notion of a zone-edge solution in which the phonon component carries all the crystal momentum.
More generally, however, the wave vector of the plane wave component is not given strictly by $q = \pi$, but by $q = \kappa$, still reflecting a state in which the phonon component carries all the crystal momentum.
The latter property suggests that outer-zone states on the weak coupling side of this transition are not simply large polaron states as we might otherwise characterize them, but mixtures of large polarons essentially at rest and ``unbound'' free phonons.

\begin{figure}[htb]
\begin{center}
\leavevmode
\epsfxsize = 3.2in
\epsffile{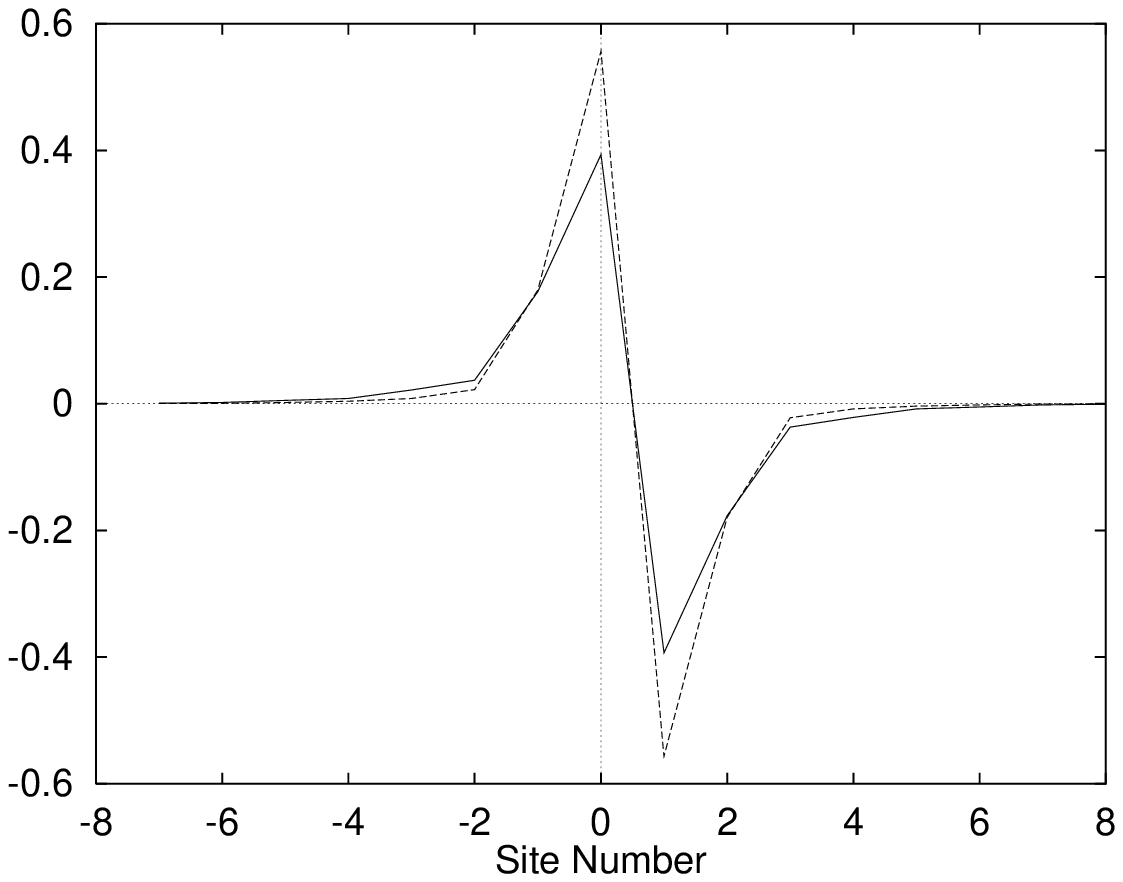}
\vspace{.1in}
\leavevmode
\epsfxsize = 3.2in
\epsffile{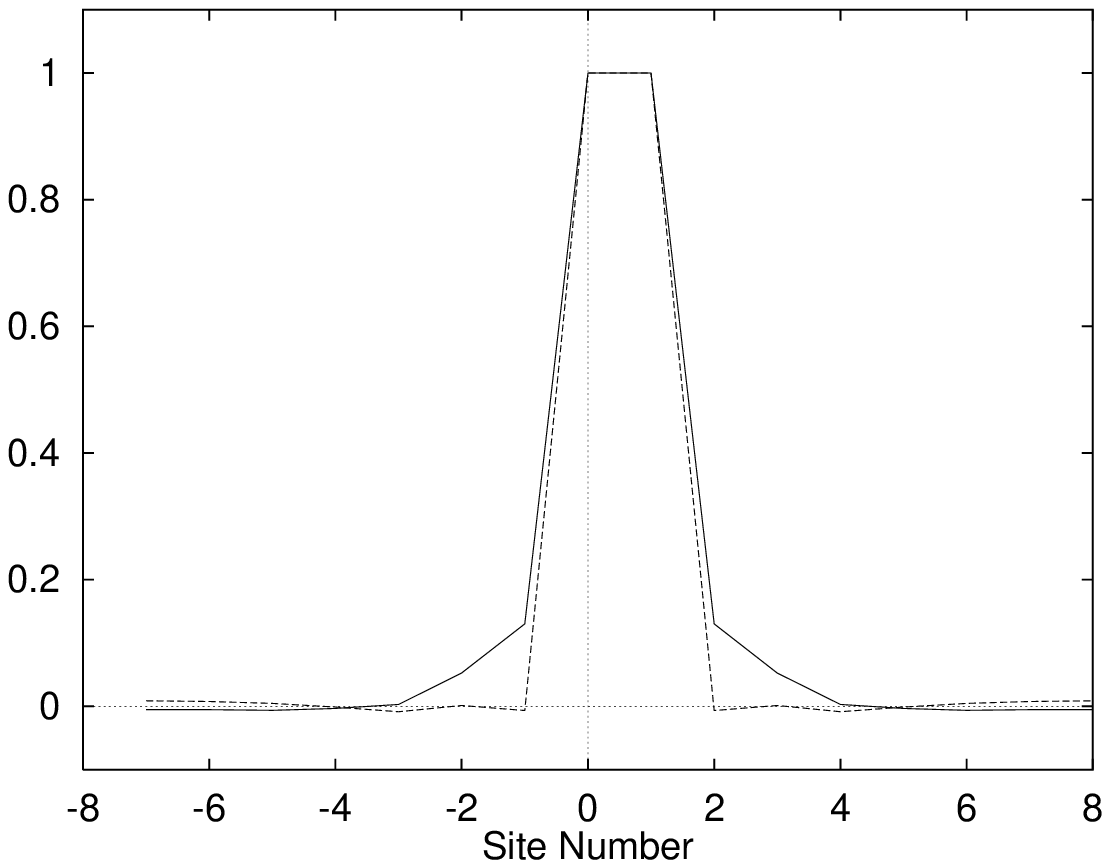}
\end{center}
\caption{
$\beta_{n}^{\kappa=0}$ (first panel) and $ \alpha_{n}^{\kappa=0}$ (second panel) for $( J, g, \phi ) = ( 2, 0, 1.6 )$ (solid line), and $( J, g, \phi) = ( 1, 0, 1.6)$ (dashed line).
}
\label{f4.7}
\end{figure}

\begin{figure}[htb]
\begin{center}
\leavevmode
\epsfxsize = 3.2in
\epsffile{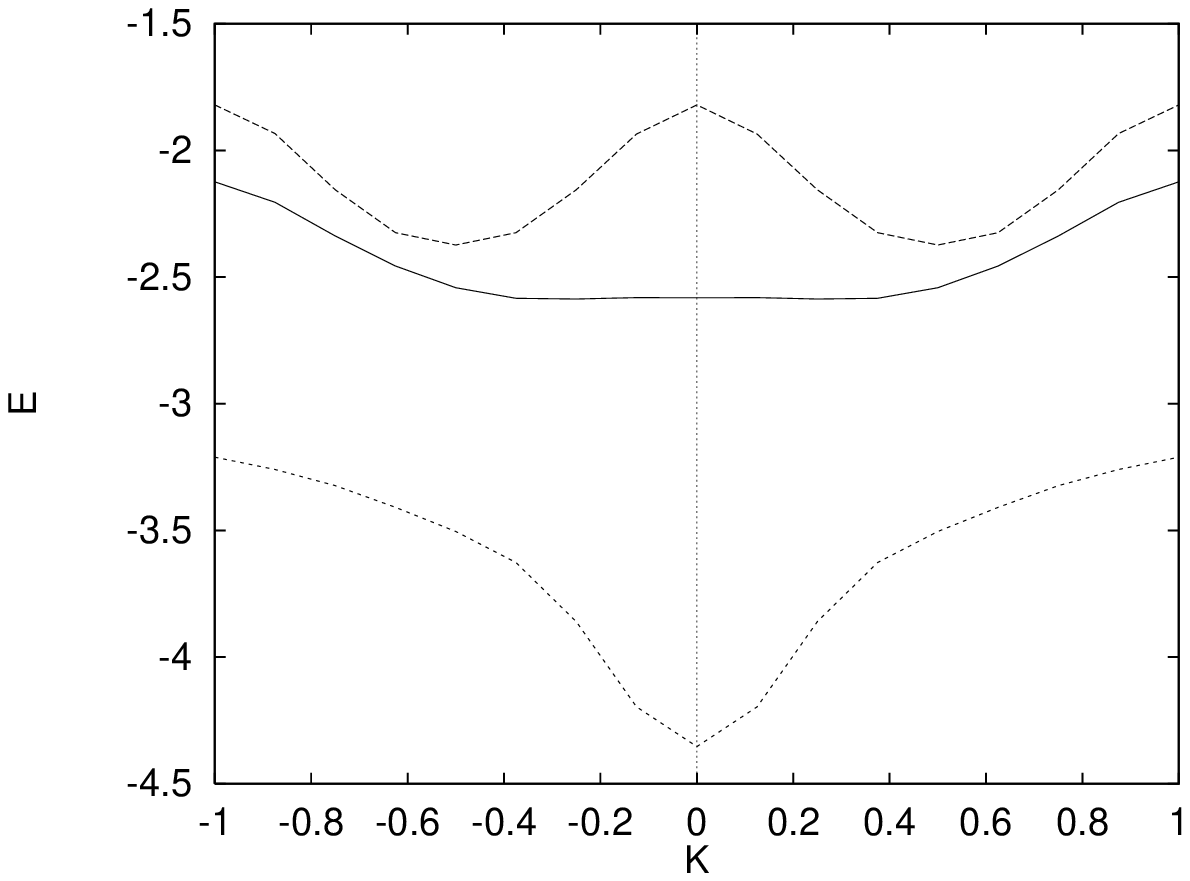}
\end{center}
\caption{
Polaron bands for $g=0, \phi=1.6$, and $J=0$ (top, dashed), $J=1$ (middle, solid), and $J=2$ (bottom, dotted), calculated from Toyozawa's Ansatz.
}
\label{f4.11}
\end{figure}

The changes in polaron structure that occur as $J$ is varied from small through large values at (relatively) fixed $\phi$, can be seen in Figures \ref{f4.3}a and \ref{f4.7}; corresponding changes in the polaron energy band are illustrated in Figure \ref{f4.11}.
The particular value of $\phi$ used in this illustration was chosen to be significant on an absolute scale, but to fall below both the zone-center and zone-edge transitions; this corresponds to a large polaron region in the sense that applies to nonlocal coupling.

At $(J,g,\phi ) = (0,0,1.6)$, the structure of the polaron state is essentially identical to that shown in Fig~\ref{f4.3}a, with slightly larger amplitudes and weaker decays due to the somewhat larger value of $\phi$.
The energy band at this point (see Figure \ref{f4.11}) has a strongly bimodal variation characterized by a negative effective mass at the zone center.
These qualitative characteristics of the polaron state and energy band are characteristic of the small $J$ regime.
At $(1,0,1.6)$, on the other hand, the exciton density is essentially completely localized on the two central sites, essentially vanishing elsewhere.
This coincides with a critcal flattening of the polaron energy band at the zone center resulting in a loss of the bimodal band structure and a divergence of the effective mass through negative values.
With further increases in $J$, e.g. to $(2,0,1.6)$, the exciton amplitudes again spread in space with a structure dominated by $J$, and the energy band is unimodal with a finite, positive effective mass.

\section{$J$ , \lowercase{$g$} and $\phi$}

\begin{figure}[htb]
\begin{center}
\leavevmode
\epsfxsize = 3.2in
\epsffile{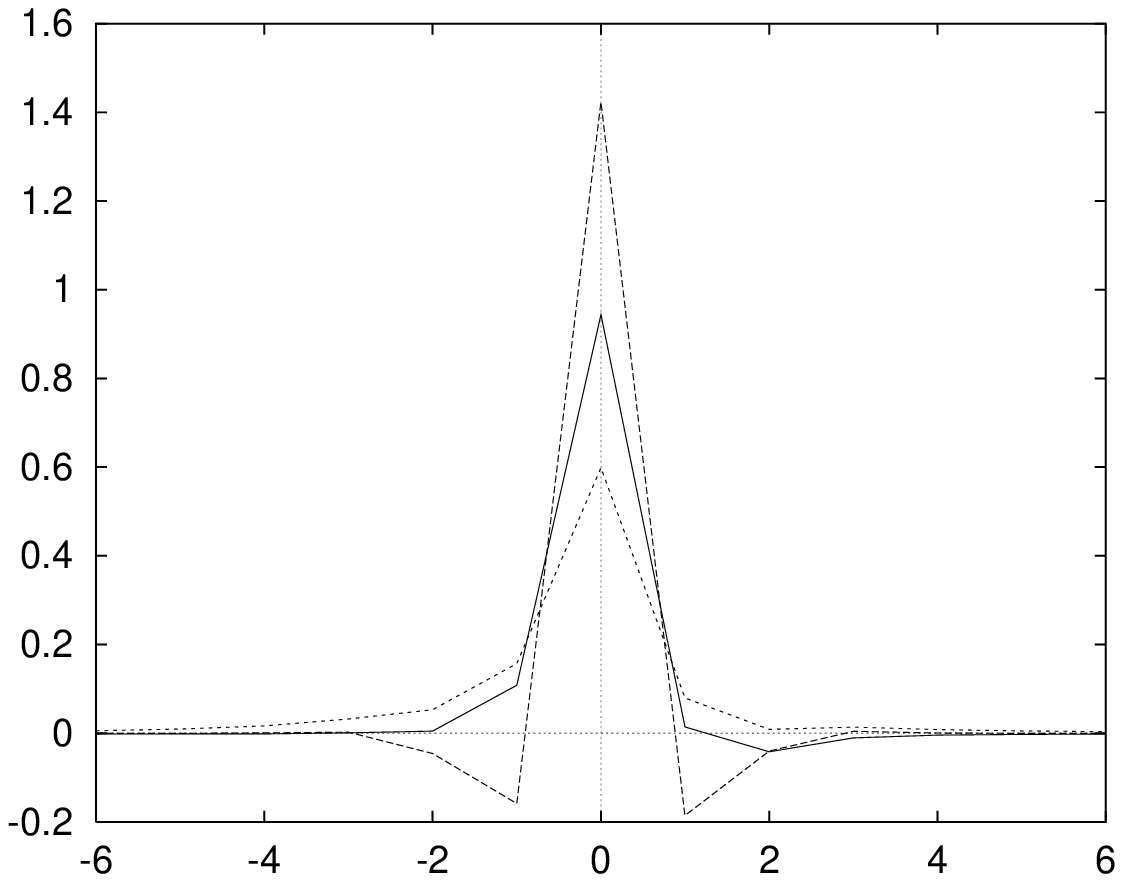}
\vspace{.1in}
\leavevmode
\epsfxsize = 3.2in
\epsffile{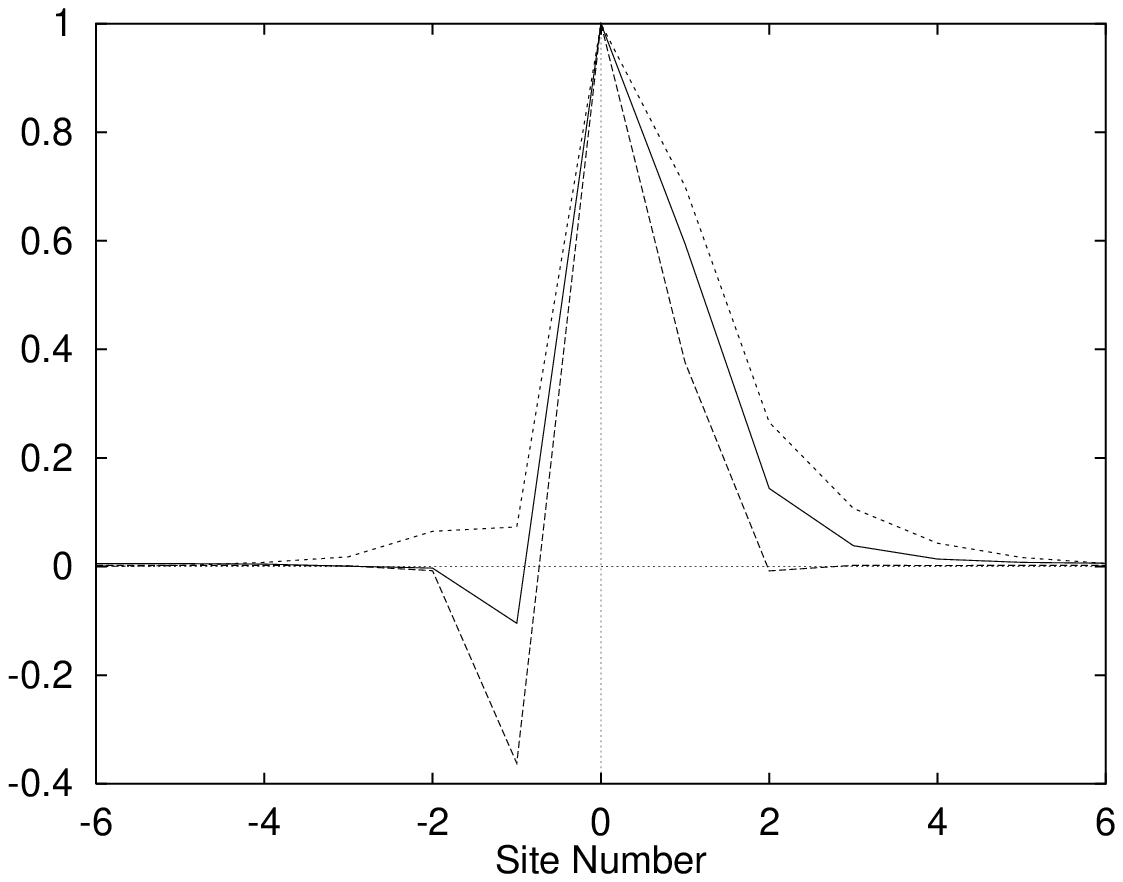}
\end{center}
\caption{
Variational parameters for the $\kappa~=~0$ state calculated from Toyozawa's Ansatz.
First panel: the phonon displacement $\beta_n^{\kappa=0}$ for $g=\phi=1$, $J=0$ (dashed line), $J=1$ (solid line) and $J=3$ (dotted line);
Second panel: the exciton amplitude $\alpha_n^{\kappa=0}$ for $g=\phi=1$, $J=0$ (dashed line), $J=1$ (solid line) and $J=3$ (dotted line).
}
\label{f4.12}
\end{figure}

\begin{figure}[htb]
\begin{center}
\leavevmode
\epsfxsize = 3.2in
\epsffile{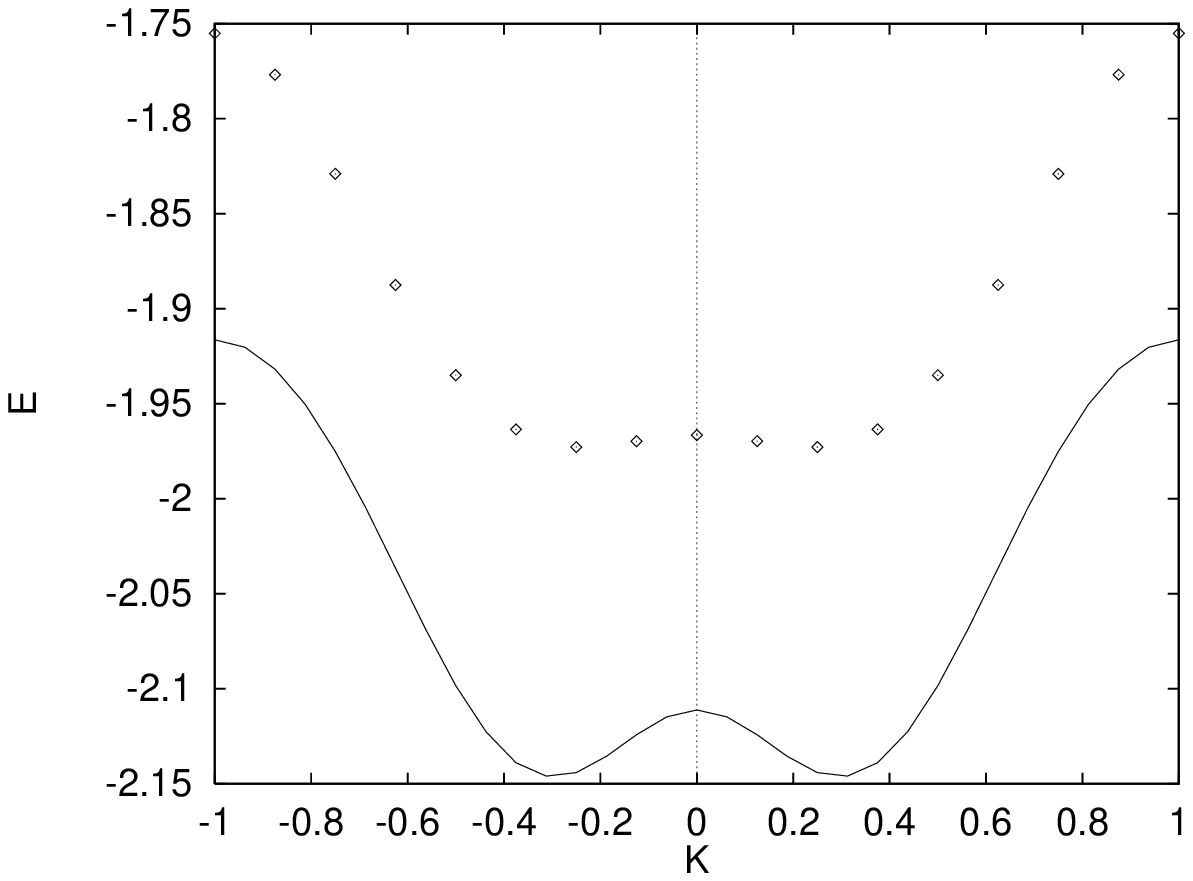}
\end{center}
\caption{
The polaron bands for $(J, g, \phi ) = ( 0.5, 1, 1 )$ at $T=0$, calculated from Toyozawa's Ansatz (solid line) and the Munn-Silbey approach \protect\cite{Zhao94a}\protect (points).
}
\label{f4.9}
\end{figure}

Allowing the simultaneous action of local and nonlocal coupling in the presence of finite transfer integrals should lead to polaron structures and energy bands that blend the qualities seen in Figures \ref{f4.3} and \ref{f4.8} ($J=0$; $g, \phi$ finite) and Figures \ref{f4.6}, \ref{f4.10}, \ref{f4.7}, and \ref{f4.11} ($g=0$; $J, \phi$ finite).
Examples of such mixed results for general $(J,g,\phi )$ are shown in Figures ~\ref{f4.12} and \ref{f4.9}.

Figure~\ref{f4.12} shows how the exciton amplitudes and the phonon displacements vary with $J$ for fixed, moderate values of both the local and nonlocal coupling coefficients.

In the absence of the transfer integral ($J=0$), the phonon displacement is mostly localized on a single site due to the dominance of local exciton-phonon coupling in this case.
The effect of nonlocal coupling shows both in the nontrivial spread of the exciton and phonon amplitudes in space (phonon-assisted transport) and in the alternations in sign that persist, though weakened by competition with local coupling.
Increasing the transfer integral $J$ results in increased spreading that further smoothes both the exciton and phonon amplitudes.
At $J=3$, all amplitudes are positive, and the influence of nonlocal exciton-phonon coupling is evident only in the asymmetry of the exciton distribution and its lattice distortion.

In the absence of the transfer integral, Toyozawa's Ansatz and the Munn-Silbey approach yield very similar polaron energy bands, though the variational approach based on Toyozawa's Ansatz is hereby established as quantitatively superior.
It is well to ask, however, {\it how} superior the present method is, and under what circumstances that difference makes a practical difference.
The first {\it semi}-quantitative conclusion in this regard is that the quantitative discrepancies increase with the transfer integral, confirming one of our expectations that motivated this work; the nature of this trend can be seen in a comparison of the lower curves of Figure~\ref{f4.8} (both methods for $(0,1,1)$) with the curves of Figure~\ref{f4.9} (both methods for $(0.5,1,1)$).
Whether the discrepancies illustrated in Figure~\ref{f4.9} are significant depends on the purpose to which a band structure calculation is to be put.
The typical ``gap'' between the two results is roughly 0.18 in units of the Einstein frequency, 0.09 in units of the rigid lattice energy bandwidth, or about 0.75 in units of the resulting polaron energy band width.
For Einstein frequencies typical of molecular phonons, a shift of 0.18 is easily resolved by optical spectroscopies, so for such purposes one would certainly want to use the present method.
On the other hand, the absolute position of the energy band is of less importance to other purposes; for example, energy transport in the band regime is more sensitive to the {\it shape} of the energy band than to its position.
The gross shapes of the two energy bands in Figure~\ref{f4.9} are similar, though there are signficant differences; for example, the ``energy barrier'' separating the two finite-$\kappa$ minima is more than twice as large when computed by the present method, while the ``effective mass'' (at $\kappa = 0$ or at the finite-$\kappa$ minima) is less than half that computed by the Munn-Silbey method.

\section{Conclusions}

In this paper we have obtained variational estimates of the ground state energy bands for the Holstein Hamiltonian incorporating simultaneous local and nonlocal exciton-phonon coupling.
We have examined the interplay between the transfer integral $J$ and local and nonlocal exciton-phonon coupling, with an emphasis on nonlocal coupling effects.

The most obvious effect of nonlocal exciton-phonon coupling is a bimodal distortion of the polaron energy band that is most obvious in the limit of small exciton transfer integral and small local exciton-phonon coupling.
Moreover, since nonlocal coupling is a transfer mechanism, a non-trivial polaron energy band remains even when both the transfer integral and local coupling vanish.

The polaron structure induced by local coupling is ``site centered'', while that induced by nonlocal coupling is ``bond centered''.
When local and nonlocal coupling act in concert, the microscopic forces associated with each superpose with the result that polaron structure is neither completely site-centered nor completely bond centered.
The nature of these interactions is such, however, that nonlocal coupling must be relatively strong before a bond-centered component becomes noticeable against the typically more prominent site-centered component driven by local coupling.
This latter quality is consistent with findings of other analyses of solid-state excimers, where strong nonlocal coupling was found to be an essential element in producing the dimeric lattice distortions characteristic of excimers as well as other excimeric properties.

Like polarons born of local coupling only, polarons embracing simultaneous local and nonlocal coupling experience self-trapping transitions, manifested in our present calculations as in many other approximate treatments by discontinuities in the dependence of some polaron properties on system parameters.
Unlike the traditional perspective that focusses on the change in the $\kappa = 0$ polaron state as some ``self-trapping line'' is crossed in the $J-g$ plane, our analysis shows that the traditional notion of self-trapping is a limiting one, the self-trapping line corresponding to a one-dimensional boundary of a three-dimensional volume in the $(J,g,\phi)$ space in which a more general, $\kappa$-dependent self-trapping occurs.
This more general self-trapping phenomenon is associated with polaron bands that
are large-polaron-like in the inner portion of the  Brillouin zone and small-polaron-like in the outer portion of the Brillouin zone, the traditional self-trapping corresponding to the limit in which the large-polaron-like region vanishes.
This characterization of self-trapping continues to hold in the $J-\phi$ plane, where the local coupling underlying the traditional self-trapping concept no longer exists.
In addition to generalizing the traditional self-trapping concept, we find a second type of transition phenomenon occurring near the Brillouin zone boundary.
Rather than constituting a small-to-large polaron type of transition, this phenomeon appears to reflect the binding or unbinding of a free phonon.

While conveniently described as ``transitions'', each of these phenomena are best understood as reflecting rapid but smooth changes in polaron structure that are ``marked'' by discontinuties only because of the approximate nature of our method.
For example, there is no reason to expect that the ``critical point'' of one of these transitions has any special meaning other than marking the point at which the changes in polaron structure first occur too rapidly to be accurately represented by our particular calculation method, permitting discontinuities to appear.
This quality of self-trapping transitions will receive greater attention elsewhere \cite{Zhao}.

For small exciton transfer integrals ($J << 1$), Toyozawa's Ansatz and the Munn-Silbey approach yield very similar polaron energy bands; however, these bands grow increasingly dissimilar with increasing transfer integral.
In all cases studied, we have found the variational approach based on Toyozawa's Ansatz to yield polaron energy bands lower than those of the Munn-Silbey method at all $\kappa$, establishing the variational energy bands as the quantitatively superior results.

Without generalization, the methods used in this paper apply to the limit of zero temperatures.
The Munn-Sibley method, on the other hand, was constructed with the aim of maintaining a sound perturbation theory at finite temperatures.
Considering the quality of the comparison between the results of our present method and those of the Munn-Silbey method at zero temperature where the latter is {\it not} at its best, we would speculate that a direct, numerical implementation of the Munn-Silbey method would provide a sound approach at finite temperatures, provided that exciton transfer integrals are small relative to both local and nonlocal coupling.
The variational approach we have employed here needs to be generalized in order to reach more reliable conclusions about the finite-temperature scenario \cite{Emin73,Gosar75,Yarkony76}.

\section*{Acknowledgments}

One of the authors (Y.Z.) would like to thank Prof. Yu Lu for 
encouragements.

\pagebreak

\noindent
FIGURE LIST

\vspace{0.25in}
\noindent
Figure 1.
The phonon displacement factor $\xi^\kappa_q$ calculated from the small polaron Ansatz for $( J, g, \phi ) = ( 0, 1, 0.1 )$.
The variation in $\xi_q^\kappa$ is small in this case because the nonlocal coupling strength $\phi$ is weak relative to the local coupling strength $g$.

\vspace{0.25in}
\noindent
Figure 2.
Comparison of the polaron bands between the Munn-Silbey approach \cite{Zhao94a} (points) and the small polaron Ansatz (solid line) for $( J, g, \phi ) = ( 0, 1, 1,)$ at $T=0$.
The small polaron Ansatz gives a higher ground state energy and a larger bimodal variation in the polaron band.

\vspace{0.25in}
\noindent
Figure 3.
Variational parameters for the $\kappa~=~0$ state calculated from Toyozawa's Ansatz.
Second panel: the exciton amplitude $\alpha_n^{\kappa=0}$ (dashed line) and the phonon displacement $\beta_n^{\kappa=0}$ (solid line) for $( J, g, \phi ) = ( 0, 0, 1 )$;
First panel: the exciton amplitude $\alpha_n^{\kappa=0}$ (dashed line) and the phonon displacement $\beta_n^{\kappa=0}$ (solid line) for $( J, g. \phi ) = ( 0, 0.3, 1 )$.

\vspace{0.25in}
\noindent
Figure 4.
Comparison between the polaron band calculated from the Munn-Silbey approach \protect\cite{Zhao94a}\protect at $T=0$, (diamonds, $( 0, 1, 1 )$, crosses, $( 0, 0.03, 1)$ and that from Toyozawa's Ansatz (solid line, $( 0, 1, 1 )$, dashed line, $( 0, 0.03, 1)$.

\vspace{0.25in}
\noindent
Figure 5.
Phase diagram on the $J - \phi$ plane for Toyozawa's Ansatz.
The two wedges correspond to the discontinuity near the Brillouin zone center (upper wedge) and that near the Brillouin zone boundary (lower wedge), respectively.

\vspace{0.25in}
\noindent
Figure 6.
Two types of convergent solutions for $Re[ \beta^{\kappa=0}_n] $ (first panel) and $ Re[ \alpha^{\kappa=0}_n ]$ (second panel).
$( J, g, \phi ) = ( 6, 0, 4 )$.
The solid line is obtained when this point is approached from the strong coupling regime, and the dashed line is obtained when this point is approached from the weak coupling regime.

\vspace{0.25in}
\noindent
Figure 7.
Two types of convergent solutions for $Re[ \beta ^{\kappa=\pi}_n]$ (first panel) and $ Re[ \alpha^{\kappa=\pi}_n ]$ (second panel).
$( J, g, \phi ) = ( 6, 0, 3.2 )$.
The solid line is obtained when this point is approached from the strong coupling regime, and the dashed line is obtained when this point is approached from the weak coupling regime.
In the second panel, the irregular tails of the dashed line are caused by numerical difficulties in the weak coupling regime.

\vspace{0.25in}
\noindent
Figure 8.
$\beta_{n}^{\kappa=0}$ (first panel) and $ \alpha_{n}^{\kappa=0}$ (second panel) for $( J, g, \phi ) = ( 2, 0, 1.6 )$ (solid line), and $( J, g, \phi) = ( 1, 0, 1.6)$ (dashed line).

\vspace{0.25in}
\noindent
Figure 9.
Polaron bands for $g=0, \phi=1.6$, and $J=0$ (top, dashed), $J=1$ (middle, solid), and $J=2$ (bottom, dotted), calculated from Toyozawa's Ansatz.

\vspace{0.25in}
\noindent
Figure 10.
Variational parameters for the $\kappa~=~0$ state calculated from Toyozawa's Ansatz.
First panel: the phonon displacement $\beta_n^{\kappa=0}$ for $g=\phi=1$, $J=0$ (dashed line), $J=1$ (solid line) and $J=3$ (dotted line);
Second panel: the exciton amplitude $\alpha_n^{\kappa=0}$ for $g=\phi=1$, $J=0$ (dashed line), $J=1$ (solid line) and $J=3$ (dotted line).

\vspace{0.25in}
\noindent
Figure 11.
The polaron bands for $(J, g, \phi ) = ( 0.5, 1, 1 )$ at $T=0$, calculated from Toyozawa's Ansatz (solid line) and the Munn-Silbey approach \protect\cite{Zhao94a}\protect (points).

\end{document}